\documentclass[prl,twocolumn,showpacs,preprintnumbers,amsmath,amssymb,superscriptaddress]{revtex4}

\usepackage{graphicx}
\usepackage{dcolumn}
\usepackage{bm}

\usepackage{color}

\begin{document}

\title{5\textit{d}-5\textit{f} Electric-multipole Transitions in Uranium Dioxide Probed  by Non-resonant Inelastic X-ray Scattering}

\author{R. Caciuffo}\affiliation{European Commission, Joint Research Centre, Institute for Transuranium Elements, Postfach 2340, D-76125 Karlsruhe, Germany}
\author{G. van der Laan}\affiliation{Diamond Light Source, Chilton, Didcot OX11 0DE, United Kingdom}
\author{L. Simonelli}\affiliation{European Synchrotron Radiation Facility, Bo\^{\i}te Postale 220 X, F-38043 Grenoble, France}
\author{T. Vitova}\affiliation{Karlsruhe Institute of Technology, Institut f\"{u}r Nukleare Entsorgung, PO Box 3640, D-76021 Karlsruhe, Germany}
\author{C. Mazzoli}\affiliation{European Synchrotron Radiation Facility, Bo\^{\i}te Postale 220 X, F-38043 Grenoble, France}
\author{M. A. Denecke}\affiliation{Karlsruhe Institute of Technology, Institut f\"{u}r Nukleare Entsorgung, PO Box 3640, D-76021 Karlsruhe, Germany}
\author{G. H. Lander}\affiliation{European Commission, Joint Research Centre, Institute for Transuranium Elements, Postfach 2340, D-76125 Karlsruhe, Germany}

\date{31 December 2009}

\begin{abstract}
Non-resonant inelastic x ray scattering (NIXS) experiments have been performed to probe the 5$d$-5$f$ electronic transitions at the uranium $O_{4,5}$ absorption edges in uranium dioxide.  For small values of the scattering vector $q$, the spectra are dominated by dipole-allowed transitions encapsulated within the giant resonance, whereas for higher values of $q$ the multipolar transitions of rank $3$ and $5$ give rise to strong and well-defined multiplet structure in the pre-edge region. The origin of the observed non-dipole multiplet structures is explained on the basis of  many-electron atomic spectral calculations. The results obtained demonstrate the high potential of NIXS as a bulk-sensitive technique for the characterization of the electronic properties of actinide materials.
\end{abstract}


\maketitle

Non-resonant inelastic x ray scattering (NIXS) is emerging as a powerful bulk-sensitive probe of the dynamical electron-density response of solids and as a new tool for the investigation of the electronic structure  in strongly-correlated systems \cite{macrander96,gurtubay05,larson07,haverkort07,veenendaal08}. The sharp non-dipolar pre-threshold features that dominate the NIXS energy-loss spectrum at large momentum transfer \cite{larson07} can be used to study atomic-environment, valence, and hybridization effects \cite{gordon08} and give information on states with different symmetries to those of electric-dipole  spectra obtained by x ray absorption spectroscopy (XAS). No NIXS experiments have been performed so far on actinide solids, and the main aim of this work has been to explore the potential of NIXS for studying the nature of the 5$f$ electron shell through the observation of  multipole transitions  from core $5d$ to valence 5$f$ states. The actinide series is still only modestly understood. Several fundamental questions remain unanswered, including the number of electrons in valence states, the angular momentum coupling, and the character of the bonding. These factors have a huge influence in driving the peculiar magnetic and electronic properties of actinide metals \cite{moore09}, alloys, and compounds \cite{santini09}, and the development of novel spectroscopic techniques sensitive to the ground-state properties of these systems is of high importance.

Compared to resonant inelastic x ray scattering (RIXS) and electron energy loss spectroscopy (EELS), the NIXS technique is penalized by a lower intensity, but it has the advantage that it permits a straightforward and quantitative interpretation of the spectra. Element specificity is provided by the energy positions of the core to valence transitions. Moreover, even when interrogating low-energy edges, NIXS remains a bulk-sensitive probe since the sampling depth is largely determined by the incident photon energy. The high penetration depth of hard X-rays is also advantageous if the sample must be inside an outer container, e.g., if thermodynamical parameters such as temperature and pressure must be varied, or when the sample must be encapsulated for safety reasons.

Here we report the results of NIXS studies on UO$_{2}$, a widely investigated semiconducting compound exhibiting long-range order of magnetic-dipole and electric-quadrupole moments at low temperature \cite{caciuffo99,wilkins06,santini09}. The U$^{4+}$ ions have a 5$f^{2}$ configuration with a $^{3}H_{4}$ ground state, with localized  $5f$ states in the 5 eV gap between the O 2$p$ band and empty U 6$d$ band, with a  $5f^{2} \to 5f^{1}6sd$ excitation energy of 2-3 eV \cite{schoenes80}.

The experiment has been performed using the ID16 inverse-geometry, multiple-analyzer-crystal spectrometer at the European Synchrotron Radiation Facility in Grenoble, France \cite{verbeni09}. The beam generated by three consecutive undulators was monochromatized  by a Si(111) double-crystal, and horizontally focussed by a Rh-coated mirror. The beam size at the sample position was 0.3$\times$1.3 mm$^{2}$ (horizontal $\times$ vertical). A set of nine
spherically-bent Si(660) analyzer crystals with 1 m bending radius, horizontal scattering geometry and vertical Rowland circles, provided a bandwidth of 1.3 eV at a final photon energy $E_f$ of 9.689 keV, and an intensity of 7$\times$10$^{13}$ photons/s for a 25 $\mu$ rad vertical divergence of the undulator radiation. The Bragg angle of the analyzers was fixed at 88.5$^\circ$, and the geometry chosen in order to measure five different momentum-transfer values simultaneously, using a position sensitive detector  based on a 256$\times$256 photon counting pixel array.
For the measurements we used a UO$_{2}$ single crystal with an external surface perpendicular to a [111] direction. A He-flow cryo-cooler gave us the possibility to cool the sample to a base temperature of 10 K.
Data have been collected in symmetric reflection geometry by scanning the incident energy $E_{i} = E_{f} + \hbar\omega$ at fixed final energy $E_{f}$, covering the $\hbar\omega$ energy-loss interval of the uranium $O_{5}$  (5$d_{5/2}\rightarrow$ 5$f$) and $O_{4}$  (5$d_{3/2}\rightarrow$ 5$f$) absorption edges. Typical counting time was six hours per spectrum.
Cross sections are particularly large between shells of the same principal quantum number, such as  for the $5d \to 5f$ transition. We also performed measurements across the uranium $N_{4,5}$   (4$d \rightarrow$ 5$f$) absorption edges, but we failed to observe NIXS signals at these edges due to the much lower cross section.

The double differential cross section for NIXS events can be expressed within the first Born approximation and is proportional to a dynamic structure factor $S(\textbf{q},\omega)$ related to the complex dielectric-response function $\varepsilon(\textbf{q},\omega)$ by the fluctuation dissipation theorem \cite{nozieres59,schulke86}, $S(\textbf{q},\omega) \propto q^{2} \mathrm{Im}[ \varepsilon(\textbf{q},\omega)^{-1}]$. The non-resonance radiation-matter interaction is dominated by a term proportional to the square of the vector potential $\textbf{A}$ and is given by \cite{schulke07}
\begin{equation}
\label{dsdw}
\frac{d^{2}\sigma}{d\Omega d\omega_{f}}  \! =  \! r_{0}^{2} \frac{ E_{f}}{E_{i}}
\sum_{\psi_{f}} |\epsilon_{i} \cdot \epsilon_{f}^\star|^2
\left|  \langle \psi_{f} \right| \mathrm{e}^{\mathrm{i} \textbf{q} \cdot \textbf{r}} | \psi_{i} \rangle|^{2}
 \delta (E_{\psi_{i}}-E_{\psi_{f}}+\hbar \omega),
\end{equation}
\noindent
where $\textbf{q} = \textbf{k}_{i}-\textbf{k}_{f}$ is the scattering vector,  $\epsilon_{i, f}$ and $\textbf{k}_{i, f}$ are respectively the polarization and wave vectors of the incident and scattered radiation. By expanding the transition operator
$\mathrm{e}^{\mathrm{i} \textbf{q} \cdot \textbf{r}}$ in terms of spherical harmonics Eq.~(\ref{dsdw}) can be written as \cite{haverkort07,gordon08}
\begin{eqnarray}
\label{sphar}
\nonumber
\frac{d^{2}\sigma}{d\Omega d\omega_{f}}  \propto&&
\sum_{\psi_{f}} \, \sum_{k=0}^{\infty} \, \sum_{m=-k}^{k}D_{k,m}
|\langle \psi_{f}(r)|j_{k}(qr)|\psi_{i}(r) \rangle|^{2}\\
 && \times \,\, \delta(E_{\psi_{i}}-E_{\psi_{f}}+\hbar \omega),
\end{eqnarray}
where the factors $D_{k,m}$ describe the angular dependence of the signal and $j_{k}(qr)$ are spherical Bessel functions of rank $k$.
The multipole moments $k$  for the $\ell \to \ell'$ shell transition are limited by the triangular condition, $| \ell-\ell' | \leq k \leq \ell+\ell' $, and the parity rule, $\ell+\ell'+k$ = even. Thus for  $d \to f$ transitions, only  $k$ = 1 (dipole), $k$ = 3 (octupole), and $k$ = 5 (triakontadipole) transitions are allowed.
The dipole transition dominates the response of the system for low values of $q$, whereas at  higher $q$ values the higher multipole transitions become important. The different $2^k$-pole transitions excite the system to different subsets of different symmetries and energies within the final state configuration.

\begin{figure}        
\includegraphics[width=7.4cm]{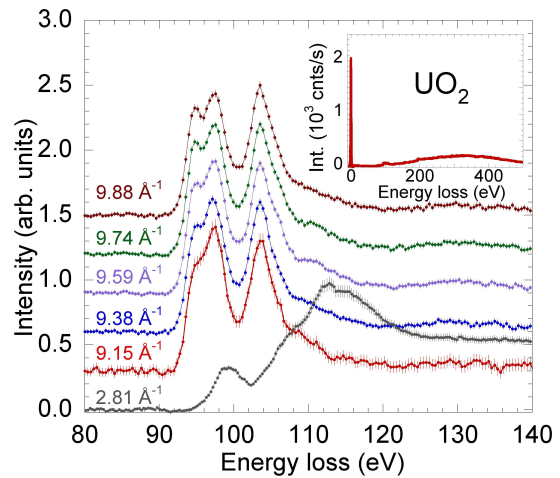}
\caption{(Color online). NIXS spectra measured for UO$_{2}$ at the uranium O$_{4,5}$ edges with a fixed final energy of 9.689 keV, at different values of the scattering vector $q$. The inset shows the spectrum measured in an extended energy range, including the elastic and Compton peaks. The NIXS signal has been normalized to the peak  at 104 eV in the high-q spectra.
\label{UO2}}
\end{figure}

The NIXS spectra obtained for UO$_{2}$ at room temperature are shown in Fig.~\ref{UO2}. The spectrum measured for $q$ = 9.6~\AA$^{-1}$ over the whole energy-loss range investigated is shown in the inset. The NIXS signal has been obtained by fitting the broad Compton peak to a polynomial function and by subtracting it from the total measured intensity. The data at different $q$ values have been normalized to the peak intensity of the feature at about 104 eV.
The  dipole spectrum observed at $q$ = 2.81~\AA$^{-1}$ shows the typical, ill-defined giant-resonance shape together with a prepeak around 99 eV, in good agreement with earlier XAS \cite{kalkowski87} and EELS  \cite{moore07,butterfield08} measurements.
Increasing $q$ above 9~\AA$^{-1}$, dipole transitions become negligible while higher multipole transitions appear at lower energies. An asymmetric peak with a maximum at 103.6~eV and a clearly-resolved double-peak feature with maxima at 94.9 and 97.3~eV characterize the spectrum at $q$ = 9.88~\AA$^{-1}$. NIXS spectra for UO$_{2}$ have also been measured at 10 K, in the ordered antiferromagnetic-antiferroelectric quadrupole phase. A comparison with the room temperature spectra does not reveal differences beyond the statistical error.

\begin{figure}       
\includegraphics[width=7.3cm]{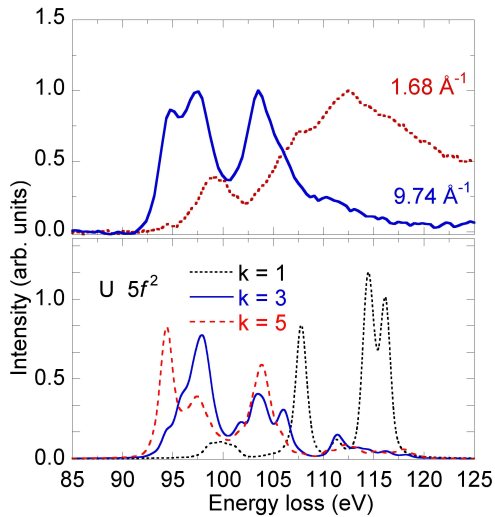}
\caption{(Color online). Top panel: Experimental NIXS spectra for UO$_{2}$ measured at 1.68~\AA$^{-1}$ (red dashed line) and 9.74~\AA$^{-1}$ (blue solid line) compared to results of atomic multiplet calculations for  the U $5f^{2}$ configuration (lower panel). Calculated contributions to the cross section from  dipole ($k$ = 1), octupole ($k$ = 3), and triakontadipole ($k$ = 5) transitions are shown separately.
\label{theory}}
\end{figure}

Many-electron atomic spectral calculations enable us to identify the origin of the observed multipole transitions.
The initial- and final-state wave functions have been calculated in intermediate coupling using Cowan's atomic Hartree-Fock (HF)  code with relativistic corrections \cite{cowan81}, which gives the
Slater parameters, $F^2_{ff}$ = 9.711, $F^4_{ff}$ = 6.364,  $F^6_{ff}$ = 4.677, $F^2_{df}$ = 10.652, $F^4_{df}$ = 6.850, $G^1_{df}$ = 12.555, $G^3_{df}$ = 7.768, $G^5_{df}$ = 5.544, and spin-orbit parameters, $\zeta_{5f}$ = 0.274, $\zeta_{5d}$ = 3.199 (all values in eV).
Angle-integrated electric $2^k$-pole transitions have been calculated from initial state U $5f^{2}$ to the final-state levels of the U $5d^{9}5f^{3}$ configuration in spherical symmetry, implicitly using the selection rules and symmetry constraints. The calculated line spectra have been broadened by a Lorentzian of $\Gamma =0.2$ eV and a Gaussian of $\sigma =0.4$ eV. The results are shown in Fig.~\ref{theory}, where the Slater parameters were scaled to 70\% of the HF values to account for screening effects.
The decay processes leading to the giant resonance were not included in the calculation. This affects the broadening of dipole spectrum, as will be discussed below.

The cross sections of the three different channels change  gradually as a function of $q$. At $q$ = 2.81~\AA$^{-1}$ only dipole transitions play a role, but with increasing $q$ the higher multipole transitions become prominent.
The ratio of the octupole to triakontadipole cross sections is $\sim$0.6 and  $\sim$0.85 at $q$ = 9.15 and 9.88~\AA$^{-1}$, respectively, with negligible dipole cross section.

Fig.~\ref{UO2} shows that for $q$ = 2.81~\AA$^{-1}$ the dipole-allowed peaks lie within the energy region of the giant resonance. The prepeak at $\sim$99 eV has previously been attributed to the finite spin-orbit interaction \cite{moore07,butterfield08}.
Going from $q$ = 9.15 to 9.88~\AA$^{-1}$, the experimental data in Fig.~\ref{UO2} show changes that are in agreement with the calculated spectra, noting that  with increasing $q$ the $k=$5 contribution increases with respect to the $k$ =  3.
The intensity of the low-energy peak in the doublet structure ($\sim$95 and  97 eV) increases relative to that of the high-energy peak with increasing q. Concurrently, the asymmetric peak at $\sim$104 eV increases in intensity and also shifts slightly to lower energy, confirming that it contains two different $k$ contributions. Thus we can conclude that there is a good overall agreement between experimental and calculated spectra.
In order to explain the differences between the multipole spectra and to understand the pronounced
multiplet structure we need to go into the details of the selection rules and the interplay between electrostatic and spin-orbit interactions.
In NIXS, the final-state  $J'$ and $L'$ values are restricted by $|J-k| \leq J' \leq J+k$ and $|L-k| \leq L' +\Delta S \leq L+k$, respectively, where $\Delta S$ = 0 in $LS$ coupling and $\Delta S$ = $\pm1$ in intermediate coupling.
Thus for initial state  $5f^2$ $^3H_4$, dipole transitions are allowed to final states $^3G$, $^3H$, $^3I $ with $J'$ = 3, 4, 5, where in intermediate coupling  quintet-spin states can mix in by $5d$ core spin-orbit interaction.
The selection rules have implications for the final states that can be reached.
For the transitions $f^n \to d^9f^{n+1}$ with less than half-filled shell, the maximum spin in the initial state is one less than in the final state. Since electric-multipole transitions obey the selection rule $\Delta S$ = 0, the high-spin final state is a {\it{ghost}} state, which cannot be reached from the initial state without spin-orbit interaction \cite{thole88}.
A similar situation arises for  $L'$ and $J'$. While the $f^2$ ground state has  $L$ = 5 and $J$ = 4, the final state  $d^9f^3$ contains values up to $L'=8$ and $J'=7$. High $L'$ and $J'$ values cannot be reached by dipole transitions, only by higher multipole transitions.

\begin{figure}       
\includegraphics[width=7.4cm]{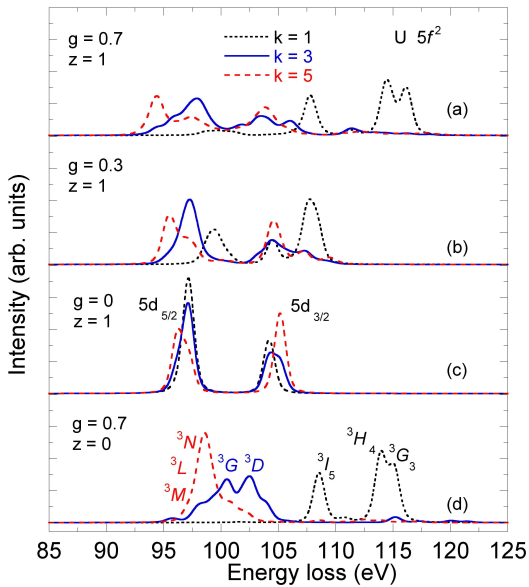}
\caption{(Color online).
Calculated  multipole spectra with $k=1$ (black dotted line), $k=3$ (blue solid line), and $k=5$ (red dashed line) for the transition U $5f^2 \to 5d^95f^3$ with different scaling factors $g$ for the Slater integrals and $z$ for the spin-orbit interaction.  (a) Realistic values: $g=0.7$, $z=1$ (intermediate coupling scheme applies), (b) strongly reduced electrostatic interactions: $g=0.3$, $z=1$, (c) only spin-orbit interaction ($jj$-coupling scheme applies), and (d) only electrostatic interactions ($LS$-coupling scheme applies). The approximate peak assignment is indicated when meaningful.  As states are strongly mixed in intermediate coupling peak assignment is less meaningful.
\label{explana}}
\end{figure}

Next, we must consider the energy distribution of the allowed final states.
The electrostatic interactions give an energy spread over the different $L'S'$ states of  about 30~eV (see Fig.~\ref{explana}).
In general,  states with higher $S'$ and $L'$ values have lower energies; however, it must be stressed that the second Hund's rule is not strictly valid  in the final state, so that states with intermediate $L'$ value can have a lower energy \cite{thole88}.
Fig.~\ref{explana}(d) shows that in $LS$ coupling, where we only take the electrostatic interactions into account, the dipole-allowed final states have much higher energies than the other states. The reason is that if the momentum transfer in the excitation process is small, the charge cloud in the final state has to be oriented in approximately the same way as in the initial state, which results in a high electrostatic energy. In contrast, higher multipole transitions allow final states where the angular distribution of the charge density is more evenly spread, thereby reducing the electrostatic energy.

Generally, $L'S'$ states mixed by spin-orbit interaction will have an energy separation of $\Delta E$ =  $\sqrt{( \Delta_{\mathrm{electrostatic}})^2 + ( \Delta_{\mathrm{spin-orbit}})^2 }
$. In the case of the dipole-allowed $L'S'$ states with triplet spin, located at higher energies, the spin-orbit coupling switches on the prepeak at $\sim$99 eV with quintet-spin character \cite{moore07}. Fig.~\ref{explana}(a-b) shows that in the dipole spectrum both peak position and prepeak intensity are most sensitive to scaling of the electrostatic interactions. On the other hand, the higher multipole allowed $L'S'$ states are closely grouped together so that they experience a strong spin-orbit interaction.  Consequently, these states will be split into two groups with an energy  separation of about  $\frac{5}{2}  \zeta_{5d} \approx 8$ eV;  one group with core hole $5d_{5/2}$ and the other with $5d_{3/2}$ character  (see Fig.~\ref{explana}). Thus the core hole spin-orbit interaction manifests itself prominently by the main energy splitting in the higher multipole spectra.
As a final point, we discuss the giant resonance appearing as a broad band around 112 eV. This feature appears clearly in the measured dipole spectrum, (Fig.~\ref{UO2}),  and is due to resonant decay into continuum states. Remarkably, no broad structures are observed in the higher multipole spectra, even though some excited states are close to the onset of the giant resonance. A simple explanation would be that the continuum states lie energetically near the dipole-allowed transitions. An alternative explanation might be that the continuum states have the same symmetry as the dipole-allowed states, and hence, interfere strongly, but have a different symmetry than the higher multipole-allowed states. Furthermore,   low energy states with high $L'$ values are often  {\it {double forbidden}}  \cite{moore07}, meaning they are not only forbidden in dipole excitation, but also do not have decay channels. This leads to long lifetimes and, hence, narrow peak widths.

In conclusion, we have demonstrated the potential of the NIXS technique for studying bulk actinide compounds. The intrinsic intensity limitation of the technique can be overcome by using an optimized spectrometer at  a third-generation synchrotron-radiation source.
The instrumental energy resolution allows us to resolve the multiplet structure. For the localized $5f$ electrons in UO$_2$ a good agreement is obtained with many-electron calculations in intermediate coupling.
Electric-multipole transitions from the ground state can reach only a limited subset of final states, thereby providing a fingerprint for the specific ground state.
These results are encouraging for perusing further NIXS measurements in a wide range of investigations on transuranium materials, with a focus on the dual localized-delocalized nature of the 5$f$ electrons and their instability induced by external pressure.

\begin{acknowledgments}

\end{acknowledgments}

\bibliography{NIXS_UO2}
\end{document}